\documentclass[letter,twocolumn]{jpsj3}

\usepackage{bm}

\begin{document}

\title{
Sum Rules and Asymptotic Behaviors for Optical Conductivity\\
of Nonequilibrium Many-Electron Systems
}

\author{
Akira \textsc{Shimizu}\thanks{E-mail address: shmz@ASone.c.u-tokyo.ac.jp}
and
Tatsuro \textsc{Yuge}$^{1}$\thanks{E-mail address: yuge@m.tohoku.ac.jp} }

\inst{
Department of Basic Science, 
University of Tokyo, 
Komaba, Meguro-ku, Tokyo 153-8902, Japan\\
$^{1}$IIAIR, Tohoku University, 
Aoba-ku, Sendai, Miyagi 980-8578, Japan
}

\def\runauthor{\sc A. Shimizu and T. Yuge}


\abst{
For many-electron systems, we consider a nonequilibrium state (NES) 
that is driven by a pump field(s), which is either an optical field 
or a longitudinal electric field. For the differential optical 
conductivity describing the differential response of the NES to 
a probe optical field, we derive exact sum rules and asymptotic 
behaviors, which open wide possibilities for experiments.  In deriving 
these results, we have also derived universal properties of general 
differential response functions of time-dependent NESs of general 
systems.
}

\kword{
sum rule, nonlinear nonequilibrium, optical properties, pump-probe
}

\maketitle

{\em Introduction -- }
The optical conductivity tensor $\sigma^{\rm eq}_{\alpha \beta}(\omega)$ 
describes the response of an equilibrium state
to a probe optical field.
It gives much information on 
electronic properties of condensed matter 
\cite{KTH,t0,t1,t4,Basov,Vescoli,Smith,Uchida,Syro,Tobe,Iwai,Tsuji}.
In particular, 
it has been shown that 
the integrals of ${\rm Re} \, \sigma^{\rm eq}_{\alpha \beta}(\omega)$ 
and $\omega {\rm Im} \, \sigma^{\rm eq}_{\alpha \beta}(\omega)$ 
over the frequency $\omega$ 
are directly related to basic properties of 
the system such as the single-particle distribution and 
band dispersion \cite{KTH,t0,t1,t4,Basov,Vescoli,Smith,Uchida,Syro,Tobe,Iwai,Tsuji}.
Such relations, called {\em sum rules},
are therefore useful for exploring electron systems 
\cite{KTH,t0,t1,t4,Basov,Vescoli},
and have been successfully utilized for analyzing 
a large variety of 
electron systems \cite{Vescoli,Smith,Uchida,Syro,Tobe,Iwai}.
However, since an equilibrium state (of each system)
is uniquely determined by a small number of parameters 
(such as temperature), the number of controllable parameters
that affect the sum (integral) values is very small.
This fact has severely limited the usage of sum rules.

This limitation can be removed by considering the optical 
conductivity of a nonequilibrium state (NES).
A NES can be created and driven by a pump field $A$, 
which is assumed to be an optical field and/or a longitudinal electric field
(generated by, say, a battery).
The response of the NES to a probe optical field is characterized by
the differential optical conductivity tensor $\sigma^{A}_{\alpha \beta}$ 
[defined by eqs.~(\ref{def:Deltaj})-(\ref{causality})]
\cite{Iwai,Tsuji,SY2010,Y2010,S2010}.
Unlike equilibrium states and $\sigma^{\rm eq}_{\alpha \beta}$, 
the NES and  $\sigma^{A}_{\alpha \beta}$ 
depend strongly on the magnitude and functional form of $A(t)$.
Therefore, 
by tuning $A(t)$ as a new controllable parameter, 
one will be able to make the sum rules for $\sigma^{A}_{\alpha \beta}$ 
much more informative
than those for $\sigma^{\rm eq}_{\alpha \beta}$.
However, 
the problem was that, until now, 
the sum rules for $\sigma^{A}_{\alpha \beta}$ were unknown.

Note that two different configurations are possible in 
experiments on $\sigma^{A}_{\alpha \beta}$: 
(i) $A(t)$ is turned off before $\bm{a}(t)$ is applied
and
(ii) $A(t)$ is present when $\bm{a}(t)$ is applied.
We here call both configurations {\em pump-probe experiments}.
In configuration (i), the NES (created by $A(t)$ beforehand) 
might sometimes be 
approximated as a quasi-equilibrium state (QES), 
and the sum rules of $\sigma^{\rm eq}_{\alpha \beta}$ are often substituted 
for those of $\sigma^{A}_{\alpha \beta}$ \cite{Iwai}.
However, in general, the transient NES is not well approximated 
as a QES, and this substitution has not been justified.
In configuration (ii), such substitution is obviously wrong because
the NES driven by $A(t)$ is far from 
quasi-equilibrium because, for example, strong mixing phenomena
such as frequency mixing take place.
Therefore, until now, 
reliable sum rules for $\sigma^{A}_{\alpha \beta}$ were unknown
in either configuration.

In this paper, 
we derive sum rules
for $\sigma^{A}_{\alpha \beta}(\omega)$
[eqs.~(\ref{result.W}) and (\ref{result.int.Im})],
and its asymptotic behaviors 
[eqs.~(\ref{result.asym.Im}) and (\ref{result.asym.Re})],
for a general class of models 
for many-electron systems.
They hold rigorously in both configurations (i) and (ii),
even when many-body interactions are strong.

{\em Differential optical conductivity of NESs -- }
Suppose that 
an optical field, described by a vector potential $\bm{A}(t)$
(in the Coulomb gauge), 
and/or
a longitudinal electric field, 
described by a scalar potential $\phi(\bm{r}, t)$,
is applied to an 
electron system.
Since $\bm{A}$ and $\phi$ induce 
optical excitation and electrical conduction, respectively,
the system becomes a NES, 
whose density operator is denoted by $\hat{\rho}^{A}(t)$.
We therefore call $A \equiv (\bm{A}, \phi)$
the {\em pump field}.
It can be strong such that 
{\em perturbation expansion in powers of $A$ breaks 
down} \cite{SY2010,Y2010,S2010}.
Furthermore, 
we do {\em not} assume any specific functional form
(such as periodicity)
for the time dependence of $A$.

One can study properties of a NES created by $A$ by 
measuring the response to another optical field $\bm{a}(t)$,
which we call a {\em probe field}.
It brings the system into another NES, $\hat{\rho}^{A+\bm{a}}(t)$.
We are interested in the change, induced by $\bm{a}(t)$,
in the current density $\bm{j}$,
\begin{equation}
\Delta \bm{j}(t) 
\equiv
\langle \hat{\bm{j}} \rangle_t^{A+\bm{a}} 
-
\langle \hat{\bm{j}} \rangle_t^{A},
\label{def:Deltaj}\end{equation}
where
$
\langle \cdot \rangle_t^{A+\bm{a}} 
\equiv {\rm Tr}[ \hat{\rho}^{A+\bm{a}}(t) \, \cdot \, ]
$
and
$
\langle \cdot \rangle_t^{A}
\equiv {\rm Tr}[ \hat{\rho}^{A}(t) \, \cdot \, ]
$.
When $\bm{a}(t)$ is weak,
$\Delta \bm{j}(t)$
is well described in terms of 
the {\em differential optical conductivity tensor} 
$\sigma_{\alpha \beta}^{A}$ as
\begin{equation}
\Delta j_\alpha(t) 
=
\sum_{\beta}
\int_{-\infty}^t \!\!\!\!\! 
\sigma_{\alpha \beta}^{A} (t - t'; t)
f_{\beta} (t') \, dt'
+o(f).
\label{eq:linear.rel}
\end{equation}
Here,
$\bm{f}(t) = - \dot{\bm{a}}(t)$
is the probe electric field,
and 
$\alpha, \beta=x,y,z$.
Since the NES 
varies as a function of time, 
so does $\sigma_{\alpha \beta}^{A}$.
That is, $\sigma_{\alpha \beta}^{A}$ depends
not only on the time delay $\tau \equiv t-t'$ but also on $t$.
Furthermore, as eq.~(\ref{eq:sigma}) shows, 
$\sigma_{\alpha \beta}^{A}$ is generally a nonlinear functional of ${A}$
[while it is independent of $\bm{a}$]. 
Throughout this paper, 
the superscript $A$, such as those
in $\sigma_{\alpha \beta}^{A}$ and $\langle \cdot \rangle_t^{A}$,
denotes such a functional dependence.
Equations (\ref{def:Deltaj}) and (\ref{eq:linear.rel}) 
and the causality,
\begin{equation}
\sigma_{\alpha \beta}^{A} (\tau; t) = 0
\quad \mbox{for $\tau < 0$},
\label{causality}
\end{equation}
define the differential optical conductivity tensor 
of the NES driven by $A$.
It contains much more information than 
that of equilibrium states, 
$\sigma_{\alpha \beta}^{\rm eq} (\omega)$,
as we will discuss later.

Experimentally, 
$\bm{a}(t)$ is usually taken as monochromatic,
and thus $\bm{f}(t) = \bm{f} e^{- i \omega t} + c.c.$
Then, eq.~(\ref{eq:linear.rel}) reads
\begin{equation}
\Delta j_\alpha(t) 
=
\sum_{\beta}
\sigma_{\alpha \beta}^{A} (\omega; t)
f_\beta e^{- i \omega t} 
+ c.c.
+o(f),
\label{eq:linear.rel.FT}
\end{equation}
where 
$ 
\sigma_{\alpha \beta}^{A} (\omega; t)
\equiv
\int_{-\infty}^\infty 
\sigma_{\alpha \beta}^{A} (\tau; t) e^{i \omega \tau} d \tau
$ 
is the Fourier transform (FT) with respect to the time delay 
$\tau$ \cite{periodic.pump}.
One can measure 
$\sigma_{\alpha \beta}^{A} (\omega; t)$ directly 
by such experiments using eq.~(\ref{eq:measure.sigma}).
Since $\sigma_{\alpha \beta}^{A} (\tau; t)$ is real, 
${\rm Re} \, \sigma_{\alpha \beta}^{A} (\omega; t)$
and 
${\rm Im} \, \sigma_{\alpha \beta}^{A} (\omega; t)$
are even and odd functions of $\omega$, respectively.
We study sum rules for them.
For example, we consider 
\begin{equation}
W_{\alpha \beta}^{A}(t) \equiv
\int_{-\infty}^\infty 
{\rm Re} \, \sigma_{\alpha \beta}^{A} (\omega; t) d \omega,
\label{def:W}\end{equation}
which is called the {\em optical spectral weight}.
This quantity is of central interest in many 
theories and experiments \cite{t0,t1,t4,Basov,Vescoli,Smith,Uchida,Syro,Tobe,Iwai,Tsuji}.

{\em Model and definitions -- }
We consider a many-electron system
in the presence of electron-electron and electron-phonon interactions 
as well as random potentials.
The electrons move on a regular lattice, 
whose {\em dimensionality and symmetries are arbitrary}.

We assume that the system is described,
in the energy scale of interest, 
by the general Hamiltonian;
\begin{equation}
\hat{H}^0
=\hat{H}_{e} + \hat{H}_{ei} 
+ \hat{H}_{ee} + \hat{H}_{ep} + \hat{H}_{p}.
\end{equation}
Here, $\hat{H}_{e}$ is the kinetic-energy term of electrons;
$ 
\hat{H}_{e}
\equiv
\sum_{\bm{k},\sigma}
\varepsilon(\bm{k}) \hat{n}_{\bm{k} \sigma},
$ 
where 
$\varepsilon(\bm{k})$ denotes the energy dispersion
of the band of interest,
and 
$\hat{n}_{\bm{k} \sigma} \equiv
\hat{c}^\dagger_{\bm{k} \sigma} \hat{c}_{\bm{k} \sigma}$.
Here, $\hat{c}_{\bm{k} \sigma} \equiv \sum_{\bm{l}} 
e^{i \bm{k} \cdot \bm{l} } \hat{c}_{\bm{l} \sigma} / \sqrt{N}$,
where
$\hat{c}_{\bm{l} \sigma}$ annihilates an electron
on site $\bm{l}$ with spin $\sigma$,
and $N$ is the number of unit cells. 
$
\hat{H}_{ei} 
\equiv \sum_{\bm{l}, \sigma} u_{\bm{l}} \hat{n}_{\bm{l} \sigma}
$
is a random potential
(with a random on-site energy $u_{\bm{l}}$ and 
$
\hat{n}_{\bm{l} \sigma} \equiv 
\hat{c}^\dagger_{\bm{l} \sigma} \hat{c}_{\bm{l} \sigma}
$),
which may be produced, for example, by impurities.
Furthermore, 
$\hat{H}_{ee}$ is the sum of electron-electron interactions.
We assume that $\hat{H}_{ee}$ is a function
of $\hat{n}_{\bm{l} \sigma}$'s.
$\hat{H}_{ep}$
is the electron-phonon interaction, 
and $\hat{H}_{p}$ denotes the Hamiltonian of free phonons.
This general model includes many models such as 
the Hubbard model (for which 
$\hat{H}_{ee} = U \sum_{\bm{l}} \hat{n}_{\bm{l} \uparrow} 
\hat{n}_{\bm{l} \downarrow}$,
$\hat{H}_{ei} = \hat{H}_{ep} = \hat{H}_{p} = 0$). 
Our results 
hold irrespective of the details and magnitudes of 
$\hat{H}_{ee}, \hat{H}_{ei}$ and $\hat{H}_{ep}$.

For later use, we define 
the {\em velocity vector} and {\em inverse mass tensor} as
\[ 
v_\alpha (\bm{k}) \equiv
{1 \over \hbar} {\partial \over \partial k_\alpha} \varepsilon(\bm{k}),
\
m^{-1}_{\alpha \beta}(\bm{k}) \equiv
{1 \over \hbar^2} {\partial^2 \over \partial k_\alpha \partial k_\beta} 
\varepsilon(\bm{k}).
\] 

To consider interactions with $\bm{A}$ and $\bm{a}$,
we assume that the spatial variations of $\bm{A}$ and $\bm{a}$
can be neglected. 
This approximation is good in most experimental configurations.
The directions 
of $\bm{A}$, 
$\nabla \phi$ and $\bm{a}$ are arbitrary.
Under these conditions,
we may incorporate the interactions with $\bm{A}$ and $\bm{a}$
by the Peierls substitution, 
and the interaction with $\phi$
by the Coulomb interaction with the charge of electrons.
Then, 
the Hamiltonian in the presence of $\bm{A}, \phi$ and $\bm{a}$
is given by
\begin{eqnarray}
\hat{H}^{A+\bm{a}}
=
\sum_{\bm{k},\sigma}
\varepsilon(\bm{k} - (e/\hbar) \bm{A}(t) - (e/\hbar) \bm{a}(t) )
\hat{n}_{\bm{k} \sigma}
+\hat{H}_{ei}
&&
\nonumber \\
+
e \sum_{\bm{l}} 
\Big( \sum_{\sigma} \hat{n}_{\bm{l} \sigma} - n^{\rm bg}_{\bm{l}} \Big) 
\phi({\bm l}, t)
+\hat{H}_{ee}
+\hat{H}_{ep}
+\hat{H}_{p}.
&&
\ \
\label{def.H}
\end{eqnarray}
Here, 
$e$ is the electron charge, and
$-e n^{\rm bg}_{\bm{l}}$ is a background charge on site $\bm{l}$.
By differentiating $\hat{H}^{A+\bm{a}}$
with $\bm{A} + \bm{a}$, we obtain
the current density as
\begin{eqnarray}
\hat{j}_\alpha
\!\!\! &=& \!\!\! 
{e \over V} \sum_{\bm{k},\sigma}
v_\alpha (\bm{k} - (e/\hbar) \bm{A}(t) - (e/\hbar) \bm{a}(t))
\hat{n}_{\bm{k} \sigma}
\label{j}
\\
\!\!\! &=& \!\!\! 
\hat{j}^v_\alpha + \hat{j}^m_\alpha 
+ o(a),
\label{j=jv+jm}\end{eqnarray}
where
\begin{eqnarray}
\hat{j}^v_\alpha
\!\!\! &\equiv& \!\!\! 
{e \over V} \sum_{\bm{k},\sigma}
v_\alpha (\bm{k} - (e/\hbar) \bm{A}(t))
\hat{n}_{\bm{k} \sigma},
\label{jv}\\
\hat{j}^m_\alpha
\!\!\! &\equiv& \!\!\! 
- {e^2 \over V} \sum_{\bm{k},\sigma, \beta}
m^{-1}_{\alpha \beta}(\bm{k} - (e/\hbar) \bm{A}(t)) 
\hat{n}_{\bm{k} \sigma}
a_\beta (t).
\label{jm}\end{eqnarray}
When $\bm{A}=0$, $\hat{j}^m_\alpha$ represents the diamagnetic current
induced by $\bm{a}$ \cite{t0,t1,t4,Basov,Vescoli}.
When $\bm{A}\neq0$, 
the diamagnetic current is induced 
by both $\bm{A}$ and $\bm{a}$,
and thus is included in both 
$\hat{j}^v_\alpha$ and $\hat{j}^m_\alpha$.

Since $\hat{j}^m_\alpha$ is $O(a)$, 
$\Delta \bm{j}(t)$ defined by eq.~(\ref{def:Deltaj}) is given by
\begin{equation}
\Delta \bm{j} (t) 
=
\Delta \bm{j}^v(t) + \bm{j}^m(t) 
+o(a).
\label{eq:Dj=Djv+jm}
\end{equation}
Here, 
$
\Delta j^v_\alpha(t) 
\equiv 
\langle \hat{j}^v_\alpha \rangle_t^{A+\bm{a}}
-
\langle \hat{j}^v_\alpha \rangle_t^{A}
$
and
$
j_\alpha^m(t) 
\equiv
- \sum_\beta d^A_{\alpha \beta}(t) a_\beta (t)
$,
where
\begin{equation}
d^A_{\alpha \beta}(t)
\equiv
{e^2 \over V} \sum_{\bm{k},\sigma}
m^{-1}_{\alpha \beta}(\bm{k} - (e/\hbar) \bm{A}(t)) 
\langle \hat{n}_{\bm{k} \sigma} \rangle_t^{A}.
\label{dA}
\end{equation}
For a simple cubic lattice, for example, 
$\sum_{\alpha \beta} d^A_{\alpha \beta}(t)$ is proportional to 
the expectation value of the kinetic energy.
 
While $\bm{j}^m(t)$ responds to $\bm{a}(t)$ instantaneously, 
$\Delta \bm{j}^v(t)$ responds with a finite delay as
\begin{equation}
\Delta j^v_\alpha(t)
=
\sum_{\beta}
\int_{-\infty}^t \!\!\!\!\! 
\Phi_{\alpha \beta}^{A} (t - t'; t)
a_{\beta} (t') \, dt'
+o(a).
\label{def:Phi}
\end{equation}
Here, $\Phi_{\alpha \beta}^{A}(\tau; t)$ is the response function 
describing the differential response of $\Delta \bm{j}^v(t)$ to $\bm{a}(t)$.
We denote its FT with respect to the time delay $\tau$ by 
$\Xi_{\alpha \beta}^{A}(\omega; t)$.
Since $\bm{f}(t) = - \dot{\bm{a}}(t)$, 
eqs.~(\ref{eq:linear.rel}) and (\ref{eq:Dj=Djv+jm})-(\ref{def:Phi}) yield
the {\em differential optical conductivity tensor} as
\begin{equation}
\sigma_{\alpha \beta}^{A} (\omega; t)
=
{-i \over \omega +i0} 
\Big[
\Xi_{\alpha \beta}^{A}(\omega; t) - d^A_{\alpha \beta}(t)
\Big].
\label{eq:sigma}
\end{equation}
Both $\Xi_{\alpha \beta}^{A}$ and $d^A_{\alpha \beta}$
are nonlinear functionals of $A$,
and so is $\sigma_{\alpha \beta}^{A}$.

{\em Universal properties of response functions of time-dependent NESs -- }
To derive sum rules for $\sigma_{\alpha \beta}^{A}$, 
we note that 
$\Xi_{\alpha \beta}^{A}$ in eq.~(\ref{eq:sigma}) should satisfy
all the universal properties that were found in ref.~\citen{SY2010}
for general response functions of general systems.
Since ref.~\citen{SY2010} assumed {\em steady} NESs
 driven by a static pump field, 
we here generalize its theory 
to time-dependent NESs, which are realized, for example, by 
the application of
a time-dependent pump field.
For this general discussion, 
we omit vector and tensor indices.

We denote the pump and probe fields by $A(t)$ and $a(t)$, respectively.
In nonequilibrium statistical mechanics
(e.g., in the Kubo formula \cite{KTH} and in 
refs.~\citen{SY2010,Y2010,S2010}),
it is usually assumed (implicitly) 
that an observable of interest
is independent of $a(t)$.
However, we here consider the general case where
an observable of interest, denoted by $\hat{Q}_{a(t)}$, 
is a function of $a(t)$,
because this is the case for $\hat{j}_\alpha$ given by eq.~(\ref{j}).
Then, 
by expanding $\hat{Q}_{a(t)}$ in powers of $a(t)$, we obtain
\begin{equation}
\hat{Q}_{a(t)} 
= 
\hat{Q} + \hat{Q}_1 a(t) 
+ o(a),
\end{equation}
where $\hat{Q}$ and $\hat{Q}_1$ are
operators independent of $a(t)$.
We have obtained such an expansion in eq.~(\ref{j=jv+jm}),
where $\hat{Q} = \hat{j}^v_\alpha$
and $\hat{Q}_1 a(t) = \hat{j}^m_\alpha$.
The response to $a(t)$,
$
\Delta Q_{a(t)}
\equiv
\langle \hat{Q}_{a(t)} \rangle_t^{A+a} 
-
\langle \hat{Q}_{a(t)} \rangle_t^{A} 
$,
is therefore given by
\begin{equation}
\Delta Q_{a(t)}
=
\Delta Q(t) 
+ \langle \hat{Q}_1 \rangle_t^{A} a(t) 
+ o(a),
\label{DA.general}
\end{equation}
where
$ 
\Delta Q(t) 
\equiv
\langle Q \rangle_t^{A+a} 
-
\langle Q \rangle_t^{A}.
$ 
Since the response function of the second term on 
the right-hand side 
is simply given by 
$\langle \hat{Q}_1 \rangle_t^{A}$,
let us consider the non-trivial term $\Delta Q(t)$.
Unlike $\langle \hat{Q}_1 \rangle_t^{A}$, 
$\Delta Q(t)$ depends on $\hat{\rho}^{A+a}(t)$
(the NES in the presence of {\em both} $A$ and $a$).
We therefore have to use 
the theory of ref.~\citen{SY2010} to evaluate $\Delta Q(t)$.

When $a(t)$ is sufficiently weak,
$\Delta Q(t)$ responds to $a(t)$ linearly as 
\begin{equation}
\Delta Q(t)
=
\int_{-\infty}^t \!\!\!\!\! 
\Phi^{A} (t - t'; t)
a(t') \, dt'
+o(a).
\label{eq:linear.rel.A}
\end{equation}
This and the causality condition,
$ 
\Phi^{A}(\tau; t) = 0
$ for $\tau < 0$,
define the differential  
response function $\Phi^{A}(\tau; t)$ of the NES.
Its FT with respect to the time delay 
$\tau$ is denoted by $\Xi^{A} (\omega; t)$.
It is straightforward
to generalize the theory of ref.~\citen{SY2010}
to the case where $A$ and the NES are time-dependent.
We then obtain the following results.

The dispersion relations, such as 
\begin{equation}
{\rm Re} \, \Xi^{A} (\omega; t)
=
\int_{-\infty}^{\infty} {{\cal P} \over \omega'- \omega} 
{\rm Im} \, \Xi^{A} (\omega'; t) {d \omega' \over \pi},
\label{eq:DR1}
\end{equation}
are satisfied.
Furthermore, the sum rules
\begin{eqnarray}
&& 
\int_{-\infty}^{\infty} \!\!
{\rm Re} \, \Xi^{A} (\omega; t) {d \omega \over \pi}
=
\langle \hat{C} \rangle^A_t,
\label{sr:ReXi}\\
&& 
\int_{-\infty}^{\infty} \!\!
\left\{ 
\omega \, {\rm Im} \, \Xi^{A} (\omega; t) 
-
\langle \hat{C} \rangle^A_t
\right\}
{d \omega \over \pi}
=
\langle \hat{D} \rangle^A_t
\label{sr:ImXi}\end{eqnarray}
hold.
Here, 
$
\hat{C}
\equiv
[ \hat{R}, \hat{Q} ]/i \hbar 
$
and
$
\hat{D}
\equiv
-[ \hat{Q}, 
[\hat{R}, \hat{H}^{A}+\hat{H}'] 
]/\hbar^2 
$,
where
$\hat{R}$ denotes the operator that couples to $a(t)$ 
via the interaction term $- \hat{R} a(t)$,
$\hat{H}^{A}$ is the Hamiltonian 
of the target system in the presence of $A$
[such as eq.~(\ref{def.H}) with $\bm{a}=0$], 
and $\hat{H}'$ is the interaction
between the target system and 
other systems such as heat reservoirs and 
electric leads \cite{SY2010}.
In general, these operators (such as $\hat{Q}$ and $\hat{R}$)
are additive operators or their densities \cite{SY2010,S2010}.
Equation (\ref{sr:ImXi}) 
also gives the asymptotic behavior for large $\omega$ as
\begin{equation}
\omega \, {\rm Im} \, \Xi^{A} (\omega; t) 
\to 
\langle \hat{C} \rangle^A_t.
\label{asym.ImXi}
\end{equation}

In deriving these results following ref.~\citen{SY2010}, 
we have used the von Neumann equation 
for the density operator of a huge system,
which includes not only the target system of interest
but also environments and a source of the pump field,
as well as all interactions among them.
[Although such a huge system is analyzed, 
we have successfully derived,
as in ref.~\citen{SY2010}, the relations among quantities 
of only the target system.]
Therefore, {\em these results are rigorous and 
apply to all physical systems}, as long as 
the linear relation given by eq.~(\ref{eq:linear.rel.A}) 
holds \cite{SY2010,Y2010}.

{\em Main results -- }
Let us apply the above results to 
$\sigma_{\alpha \beta}^{A}$ of the system
described by eq.~(\ref{def.H}).
By expanding
$\hat{H}^{A+\bm{a}}$ in powers of $\bm{a}(t)$, 
we find that 
$\hat{R} = V \hat{j}^v_\beta$ for $a_\beta(t)$.
For $\Xi_{\alpha \beta}^{A}$,
which is the FT of $\Phi_{\alpha \beta}^{A}$ of eq.~(\ref{def:Phi}), 
$\hat{Q} = \hat{j}^v_\alpha$.
The sum rules for $\sigma_{\alpha \beta}^{A}$ are obtained from
the properties of $\Xi_{\alpha \beta}^{A}$ through eq.~(\ref{eq:sigma}).

For the optical spectral weight [defined by eq.~(\ref{def:W})],
eq.~(\ref{eq:DR1}) for $\omega=0$ yields\cite{similar}
\begin{equation}
W_{\alpha \beta}^{A}(t)
=
\pi d^A_{\alpha \beta}(t).
\label{result.W}\end{equation}
%
%
Note that this result relies only on 
eqs.~(\ref{eq:sigma}) and (\ref{eq:DR1}).
That is, this sum rule is derived only from 
the causality [eq.~(\ref{causality})] 
and the specific form of the current [eq.~(\ref{j=jv+jm})]: 
No other relations are necessary for deriving this sum rule.
For $\omega \, {\rm Im} \, \sigma_{\alpha \beta}^{A}$,
on the other hand, 
eqs.~(\ref{eq:sigma}) and (\ref{sr:ReXi}) yield 
the following sum rule:
\begin{equation}
\int_{-\infty}^\infty \!\!\!\!
\left\{
\omega \, {\rm Im} \, \sigma_{\alpha \beta}^{A} (\omega; t)
-
d^A_{\alpha \beta}(t)
\right\} \! d \omega
=
0.
\label{result.int.Im}\end{equation}
%
This and eq.~(\ref{asym.ImXi}), respectively, give the asymptotic behaviors 
for large $\omega$ as
\begin{eqnarray}
\omega \, {\rm Im} \, \sigma_{\alpha \beta}^{A} (\omega; t)
&\to& 
d^A_{\alpha \beta}(t),
\label{result.asym.Im}
\\
\omega^2 \, {\rm Re} \, \sigma_{\alpha \beta}^{A} (\omega; t)
&\to& 
0.
\label{result.asym.Re}
\end{eqnarray}

Equations (\ref{result.W})-(\ref{result.asym.Re})
are our main results.
They are rigorous 
(to the same degree as the Kubo formula is) within the 
general model defined by eq.~(\ref{def.H}),
even when $\bm{A}(t), \phi(t), \hat{H}_{ee}, \hat{H}_{ep}$ and $\hat{H}_{ei}$ 
are strong.
For example, 
our results hold for any possible phases of 
the system that is described by eq.~(\ref{def.H}).
That is, our results are completely valid as long as
the target system 
is well described by the Hamiltonian of eq.~(\ref{def.H}).
Conversely, if
experimental results disagree with 
our results, it means that 
the system is not described by eq.~(\ref{def.H})
(because, say, transition to another band takes place).
Such rigor seems important for the application of the
sum rules and asymptotic behaviors.

Note that 
the effects of 
$\hat{H}_{ee}, \hat{H}_{ep}, \hat{H}_{ei}$ and $\phi$ 
on the sum and 
asymptotic values appear only 
through the distribution function
$\langle \hat{n}_{\bm{k} \sigma} \rangle_t^{A}$.
In contrast, 
the effects of $\bm{A}$ on the sum and 
asymptotic values appear 
not only through $\langle \hat{n}_{\bm{k} \sigma} \rangle_t^{A}$
but also through 
$m^{-1}_{\alpha \beta}(\bm{k} - (e/\hbar) \bm{A})$.
In either case, the decoherence of electrons affects 
the sum and asymptotic values only through the broadening of 
$\langle \hat{n}_{\bm{k} \sigma} \rangle_t^{A}$.

{\em Possible applications -- }
For $\sigma_{\alpha \beta}^{\rm eq}$, 
the sum rule for 
$
W_{\alpha \beta}^{\rm eq} \equiv
\int_{-\infty}^\infty 
{\rm Re} \, \sigma_{\alpha \beta}^{\rm eq} (\omega) d \omega
$
reads
$
W_{\alpha \beta}^{\rm eq} / \pi
=
d^{\rm eq}_{\alpha \beta}
\equiv
(e^2/V) \sum_{\bm{k},\sigma}
m^{-1}_{\alpha \beta}(\bm{k}) 
\langle \hat{n}_{\bm{k} \sigma} \rangle^{\rm eq}
$ \cite{KTH,t0,t1,t4,Basov,Vescoli}.
For each system, 
$d^{\rm eq}_{\alpha \beta}$
depends only on the temperature $T$ and doping density $n_{\rm d}$.
In pump-probe experiments, in contrast,
$W_{\alpha \beta}^{A} / \pi = d_{\alpha \beta}^{A}(t)$
can be studied as a function of $T, n_{\rm d}$ and $A$.
This opens wide possibilities for studying many-electron systems.
For example, suppose that an ordered phase is realized as 
an equilibrium state.
By measuring $\sigma_{\alpha \beta}^{\rm eq}$, 
one obtains the value of $d^{\rm eq}_{\alpha \beta}$
for the ordered phase. 
Then, a static $A=(0,\phi)$ is applied to induce
a DC electric current while 
keeping $T$ equal to that for $A=0$ (by, for example, using a good heat sink).
By measuring $\sigma_{\alpha \beta}^{(0,\phi)}$ by applying $\bm{a}(t)$, 
one now obtains, 
from eq.~(\ref{result.W}) or (\ref{result.asym.Im}),
the value of 
$
d^{(0,\phi)}_{\alpha \beta}
=
(e^2/V) \sum_{\bm{k},\sigma}
m^{-1}_{\alpha \beta}(\bm{k}) 
\langle \hat{n}_{\bm{k} \sigma} \rangle^{(0,\phi)}
$
for a {\em non}-ordered phase,
because the order would be destroyed by the electric current 
if $|\nabla \phi|$ was larger than a certain value.
One thus obtains the values of $d_{\alpha \beta}$
with and without the order at the same $T$ and $n_{\rm d}$.
Alternatively, suppose that no order is present
in an equilibrium state.
Then, a coherent optical field $A=(\bm{A}(t),0)$ is applied.
This would induce an electron-hole ($eh$) correlation.
Hence, by measuring $\sigma_{\alpha \beta}^{A}$, 
one obtains the value of $d^A_{\alpha \beta}$
for the state with the $eh$ correlation.

{\em Method of measuring $\sigma_{\alpha \beta}^{A}(\omega;t)$ --- }
$\sigma_{\alpha \beta}^{A}(\omega;t)$ can be measured,
for example, by the following process.

%

Step 1: Prepare the system in some initial state at an initial time $t=0$.
Apply a pump field $A(t)$ only, and measure 
the current density $\bm{j}(t)$ continuously for a sufficiently long time. 
Then, turn off $A(t)$, and at another initial time prepare the system 
in the same initial state as that at $t=0$.
Redefine the origin of time ($t=0$) as this new initial time.
Apply the same pump field $A(t)$  again, and measure 
the current density $\bm{j}(t)$ continuously.
By repeating these procedures sufficiently many times, 
one obtains many independent records of $\bm{j}(t)$.
The average of these records gives 
$\langle \hat{\bm{j}} \rangle_t^{A}$.

Step 2: Perform the same sequence of experiments 
using the pump and probe fields 
instead of the pump field.
Here, the pump field $A(t)$ is taken to be the same as that of Step 1.
One then obtains $\langle \hat{\bm{j}} \rangle_t^{A+\bm{a}}$. 
From this and the result of Step 1, 
one obtains 
$\Delta \bm{j}(t) 
=
\langle \hat{\bm{j}} \rangle_t^{A+\bm{a}} 
-
\langle \hat{\bm{j}} \rangle_t^{A}
$.
If one takes 
the probe field as a monochromatic one, 
$\bm{f}(t) = \bm{f} e^{- i \omega t} + c.c.$, 
and if one takes $\bm{f}$ parallel to 
the $\beta$-axis (i.e., $f_\alpha = f \delta_{\alpha \beta}$), 
then eq.~(\ref{eq:linear.rel.FT}) yields
$
\Delta j_\alpha(t) 
=
\sigma_{\alpha \beta}^{A} (\omega; t)
f e^{- i \omega t} 
+ c.c. + o(f)
$.

Step 3: Perform the same sequence of experiments 
using the same pump field and another (phase shifted) probe field 
$\bm{f}'(t) = \bm{f}' e^{- i \omega t} + c.c.$,
where $\bm{f}' = i\bm{f}$.
One then obtains $\langle \hat{\bm{j}} \rangle_t^{A+\bm{a}'}$. 
From this and the result of Step 1, 
one obtains 
$\Delta \bm{j}'(t) 
\equiv
\langle \hat{\bm{j}} \rangle_t^{A+\bm{a}'} 
-
\langle \hat{\bm{j}} \rangle_t^{A}
$.
According to eq.~(\ref{eq:linear.rel.FT}),
it is expressed as
$
\Delta j_\alpha'(t) 
=
\sigma_{\alpha \beta}^{A} (\omega; t)
i f e^{- i \omega t} 
+ c.c. + o(f)
$.

From these experimental results,
one can evaluate $\sigma_{\alpha \beta}^{A} (\omega; t)$ 
using
\begin{equation}
\sigma_{\alpha \beta}^{A} (\omega; t)
=
\lim_{f \to 0}
{
\Delta j_\alpha(t) - i \Delta j_\alpha'(t)
\over
2 f e^{- i \omega t}
}.
\label{eq:measure.sigma}\end{equation}

{\em Concluding remarks --- }
Our results hold in both configurations (i) and (ii),
which were discussed in the introduction.
In configuration (i), 
eq.~(\ref{result.W}) reads
$
W_{\alpha \beta}^{A}(t)
=
(\pi e^2 /V) \sum_{\bm{k},\sigma}
m^{-1}_{\alpha \beta}(\bm{k}) 
\langle \hat{n}_{\bm{k} \sigma} \rangle_t^{A}
$.
Comparing this with 
the corresponding result for 
$\sigma_{\alpha \beta}^{\rm eq}$ \cite{KTH,t0,t1,t4,Basov,Vescoli},
$
W_{\alpha \beta}^{\rm eq}
=
(\pi e^2 /V) \sum_{\bm{k},\sigma}
m^{-1}_{\alpha \beta}(\bm{k}) 
\langle \hat{n}_{\bm{k} \sigma} \rangle^{\rm eq}
$,
we find that 
the result for $W_{\alpha \beta}^{A}(t)$
is obtained simply by replacing
the equilibrium electron distribution
$\langle \hat{n}_{\bm{k} \sigma} \rangle^{\rm eq}$
with 
the nonequilibrium one $\langle \hat{n}_{\bm{k} \sigma} \rangle_t^{A}$.
Hence,
the analysis of the pump-probe experiments 
in ref.~\citen{Iwai}, which substituted 
the sum rule of $W_{\alpha \beta}^{\rm eq}$
for that of $W_{\alpha \beta}^{A}(t)$, is now justified.
In configuration (ii), 
on the other hand, 
eq.~(\ref{result.W}) reads
$ 
W_{\alpha \beta}^{A}(t)
=
(\pi e^2 /V) \sum_{\bm{k},\sigma}
m^{-1}_{\alpha \beta}(\bm{k} - (e/\hbar) \bm{A}(t)) 
\langle \hat{n}_{\bm{k} \sigma} \rangle_t^{A}.
$ 
Since the pump field enters the inverse mass tensor, 
the simple replacement of 
$\langle \hat{n}_{\bm{k} \sigma} \rangle^{\rm eq}$
with 
$\langle \hat{n}_{\bm{k} \sigma} \rangle_t^{A}$
in the sum rule of $W_{\alpha \beta}^{\rm eq}$
does not yield the correct result.

Finally, we point out that the present results can be generalized.
Suppose that the current density takes a general form;
\[ 
\hat{j}_\alpha
=
\hat{J}^A_{\alpha}
-\sum_{\beta} \hat{D}^A_{\alpha \beta} \, a_\beta (t)
+ o(a).
\] 
Here, $\hat{J}^A_{\alpha}$ and $\hat{D}^A_{\alpha \beta}$ are
arbitrary vector and tensor operators, respectively, 
which may be functions of $A$.
[Equation (\ref{j=jv+jm}) takes this form.]
Then the sum rules eqs.~(\ref{result.W}) and (\ref{result.int.Im}) 
are respectively generalized as
\begin{eqnarray}
&&
\int_{-\infty}^\infty 
{\rm Re} \, \sigma_{\alpha \beta}^{A} (\omega; t) \, d \omega
=
\pi \langle \hat{D}^A_{\alpha \beta} \rangle_t^{A},
\\
&&
\int_{-\infty}^\infty \!\!
\left\{
\omega \, {\rm Im} \, \sigma_{\alpha \beta}^{A} (\omega; t)
-
\langle \hat{D}^A_{\alpha \beta} \rangle_t^{A}
\right\} d \omega
=0.
\end{eqnarray}
Furthermore, generalizations to the case where the probe field is a longitudinal AC electric field and to higher-order responses
[following ref.~\citen{S2010}] are straightforward.


We thank T. Oka and N. Tsuji 
for directing our attention to this 
problem and for helpful discussions.
This work was supported by KAKENHI Nos.~22540407 and 23104707, 
and by a Grant-in-Aid for the GCOE Program 
``Weaving Science Web beyond Particle-Matter Hierarchy''.


\begin{thebibliography}{99}
\bibitem{periodic.pump}
When ${A}(t)$ is monochromatic, 
one might be interested in
$\sigma_{\alpha \beta}^{A} (\omega; \nu)$,
which is the FT of $\sigma_{\alpha \beta}^{A} (\omega; t)$
with respect to $t$.
If ${A}(t)$ is sufficiently weak,
$\sigma_{\alpha \beta}^{A} (\omega; \nu)$
has a peak at $\nu = \Omega$,
where $\Omega$ is the frequency of ${A}(t)$.
As ${A}(t)$ becomes stronger,
$\sigma_{\alpha \beta}^{A} (\omega; \nu)$
has more peaks as a function of $\nu$,
until, for a strong ${A}(t)$, it has a broad spectrum
in addition to multiple peaks.
\end{thebibliography}
\end{document}